\documentclass[aps,prd,twocolumn,superscriptaddress,amsmath,amssymb]{revtex4}
\pdfoutput=1
\usepackage{graphicx}
\usepackage{color}
\usepackage{ulem}

\begin{document}
\title{Revisiting quantum decoherence for neutrino oscillations in constant matter density}

\author{J.A. Carpio}
\affiliation{Secci\'on F\'isica, Departamento de Ciencias, Pontificia Universidad Cat\'olica del Per\'u, Apartado 1761, Lima, Per\'u}
\affiliation{Department of Physics, The Pennsylvania State University, University Park, Pennsylvania 16802, USA}
\author{E. Massoni}
\author{A.M. Gago}
\affiliation{Secci\'on F\'isica, Departamento de Ciencias, Pontificia Universidad Cat\'olica del Per\'u, Apartado 1761, Lima, Per\'u}

\begin{abstract}
We re-examine the matter neutrino oscillation probabilities considering the decoherence phenomenon as a sub-leading effect.  In this paper we 
point out the relevance of having the correct interpretation of the decoherence matrix in the different quantum bases, within the framework of 
neutrino oscillation probabilities in matter. Based on this treatment we develop an analytical formula for matter neutrino oscillation  
probabilities for three generations, with a range of application up to the decoherence parameter $\Gamma \sim 10^{-23}$ GeV. We observe that, 
due to decoherence, the amplitudes of the neutrino/antineutrino oscillation probabilities increase in an energy independent way. We also find 
that decoherence can reduce the absolute value of the CP asymmetry, relative to its value at the pure oscillation case. 
As a side effect we have found a degeneracy between the decoherence parameter $\Gamma$ and the CP violation phase $\delta$. 
\end{abstract}
\maketitle

\section{INTRODUCTION}
The neutrino oscillations caused by non-zero neutrino mass is a well established phenomenon supported by numerous experimental data 
accumulated since two decades ago \cite{Fukuda01,Ahmad02,Fukuda98,Kajita16,Araki05,An12,Adamson14}.

Even though all the evidence indicates that the neutrino oscillation relies on the interference between 
different neutrino mass eigenstates, the presence of an as yet undetected sub-leading mechanism is still possible. Within the context of new 
physics, there are various candidates that can coexist with oscillations induced by mass, that we will call from now on standard 
oscillations. Among them we have neutrino decay \cite{Berryman15,Frieman88,Raghavan88,Berezhiani92a,Berezhiani92b,Berezhiani93,Barger99,
Beacom02,Joshipura02,Bandyopadhyay03,Ando04,Fogli04,Palomares05,Gonzalez08,Maltoni08,Baerwald12,
Meloni07,Das11,Dorame13,Gomes15,Abrahao15,Picoreti16,Bustamante17,Gago17,Coloma17}
, nonstandard interactions \cite{Gonzalez99,Bergmann00,Guzzo04,Gago01b,Gago02b,Ohlsson13,Esmaili13},
decoherence in oscillations \cite{Benatti00,Benatti01,Gago02a,Oliveira10,Oliveira13,Oliveira16,Lisi00,Barenboim05,Farzan08,Bakhti15,
Guzzo16,Gago01a,Morgan06,Fogli07,Balieiro16}, and other new physics effects \cite{Pantaleone93,Bustamante10,Barger00,Colladay98,Esmaili14}.

In particular, the general consequences of considering a quantum system in interaction with its environment are irreversibility and 
decoherence. The decoherence phenomenon tends to destroy the interference pattern, through the introduction of damping terms of the type 
of $e^{-\Gamma L}$ (where $\Gamma$ is a decoherence parameter and $L$ is the neutrino source-detector distance or baseline).
It is also possible for this phenomenon to modify the oscillation frequencies through the appeareance of new coherence terms. It has been 
pointed out that the source of decoherence could be originated by strings and branes \cite{Ellis, Benattistrings}, as well as quantum gravity 
effects \cite{Hawking1}. There have been 
several papers that have included dissipative effects into the neutrino system, treating this as an open quantum system, developing the 
oscillation probabilities formulae in two and three generations, for vacuum and matter
\cite{Benatti00,Benatti01,Gago02a,Oliveira10,Oliveira13,Oliveira16}. 

Decoherence has been proposed as a possible solution for experimental 
data \cite{Lisi00,Barenboim05,Farzan08,Bakhti15} and, on the other hand, constraints on decoherence parameters have been obtained from
data \cite{Gago01a,Morgan06,Fogli07,Oliveira14,Balieiro16,Guzzo16}. More specifically, the bounds at 95\%C. L.  for atmospheric neutrinos and the MINOS long baseline experiment are: $\Gamma < 4.10 \times 10^{-23}$ GeV \cite{Lisi00}  and $\Gamma < 9.11 \times 10^{-23}$ GeV \cite{Oliveira14}, respectively. Similarly, for solar neutrinos and reactor are: $\Gamma < 0.64 \times 10^{-24}$ GeV \cite{Fogli07} and $\Gamma < 6.8 \times 10^{-22}$ GeV \cite{Balieiro16}, respectively. 

The main goal of this paper is revisiting the treatment for obtaining a three neutrino oscillation formula in matter, when 
dissipative effects are included. In this framework, the neutrino Hamiltonian in matter can be written in the vacuum mass eigenstates basis 
(VMB) or in the matter mass eigenstates basis (MMB). The latter is the basis that diagonalizes the neutrino matter Hamiltonian, for a constant matter density. When solving the system in the MMB we have to check if the decoherence matrix 
we propose in this basis can be generated from a rotation of the 
corresponding one in the VMB,  where the decoherence matrix is in fact defined. This very relevant detail has been overlooked in some papers
by assuming that the decoherence matrix can be written in MMB as diagonal and energy independent. We will show that, 
in general, the latter assumptions are fulfilled only in a few cases.   

Considering this rotation appropiately, we provide a three generation analytical formula valid for a decoherence parameter $\Gamma \leq 10^{-23}$ GeV, that corresponds to an upper limit for $\Gamma L$ of $\mathcal O(10^{-1})$, with a baseline $L$ of $\mathcal O(10^3)$ km, a 
source-detector distance compatible with long baseline scenarios. We study the behaviour of these probabilities, using the DUNE baseline 
and energy range \cite{Acciarri15}, and also explore how a CP violation measurement would be affected due to the presence of decoherence.

\section{METHOD}
\subsection{General neutrino Hamiltonian}
We work with $N$ neutrino generations going through matter with constant density. 
Sterile neutrinos get an additional contribution from the neutral current matter potential that cannot be 
ignored, since the latter cannot be factorized out as a global phase as we do with the purely active
neutrinos scheme. The neutrino Hamiltonian in the VMB for a neutrino of energy $E$ can be written as:
\begin{equation}
H_V = \frac{1}{2E}\left(\Delta M^2+U^\dag \mathbb{A} U\right)
\end{equation}
Here, $\Delta M^2=\text{Diag}(0,\Delta m^2_{21},\dots,\Delta m^2_{N1})$ is the mass term and $\Delta m^2_{ij}=m^2_i-m^2_j$ is the squared
mass difference. The matter potential term is
$\mathbb{A}=\text{Diag}(A_{CC},0,0,A_{NC},\dots,A_{NC})$, where $A_{CC}=2\sqrt{2}EG_F n_e$ and $A_{NC}=\sqrt{2}EG_F n_n$ are the 
charged current (CC) and neutral current (NC) potentials. The parameters $G_F,n_e$, and $n_n$ are the Fermi coupling constant, the electron 
and neutron number densities respectively. The mixing matrix $U$ rotates the VMB mass eigenstates into the flavour eigenstates. 

The Hamiltonian $H_V$ given above is non-diagonal, but is diagonal in the MMB and can be written as:  
\begin{equation}
H_M = \text{Diag}(0,\tilde{\Delta}_{21},...,\tilde{\Delta}_{N1})
\end{equation}
where $\tilde{\Delta}_{ij}$ becomes the effective value of $\Delta_{ij}=\Delta m^2_{ij}/2E$ in matter. 

We define $U_M$ as the matrix that rotates the MMB mass eigenstates into the flavour eigenstates and can be expressed in terms of effective 
mixing angles $\tilde{\theta}_{ij}$ and effective phases $\tilde{\delta}$. We also introduce the matrix $U_T$, which relates operators 
$\hat{O}_V$ in the VMB to operators $\hat{O}_M$ in the MMB via
\begin{equation}
\hat{O}_M = U_T^\dag\hat{O}_VU_T
\label{UnitaryTransformationEq}
\end{equation}
where we have defined $U_T=U^\dag U_M$, the matrix that rotates the mass eigenstates in the MMB into the 
mass eigenstates in the VMB. 

\subsection{Density matrix formalism}
\label{DensityMatrixFormalism}
We will consider the neutrino system coupled with the environment and treated as an open quantum system. Thus, its evolution is described by the Lindblad Master Equation
\begin{equation}
\frac{d\hat{\rho}(t)}{dt} = -i[H,\hat{\rho}(t)]+D[\hat{\rho}(t)]
\label{LindbladMasterEquation}
\end{equation}
where $\hat{\rho}(t)$ is the neutrino's density matrix and $H$ is the Hamiltonian of the system. The dissipative term $D$ is written as
\begin{equation}
D[\hat{\rho}(t)] = \frac{1}{2}\sum_j \left([\hat{\mathcal{V}}_j,\hat{\rho}(t)\hat{\mathcal{V}}_j^\dag]+[\hat{\mathcal{V}}_j\hat{\rho}(t),\hat{\mathcal{V}}_j^\dag]\right)
\end{equation}
where $\{\hat{\mathcal{V}}_j\}$ is a set of dissipative operators with $j=1,2,\dots,N^2-1$ for $N$ neutrino generations. 
The presence of the operators $\hat{\mathcal{V}}_j$ causes the evolution of $\hat{\rho}$ to be non-unitary. The dissipative term must 
satisfy the requirements of complete positivity 
and a von Neumann entropy that increases with time. The second condition is achieved by requiring $\hat{\mathcal{V}}_j$ to be Hermitian 
\cite{Benatti88}. The usual approach is to rewrite Eq. \eqref{LindbladMasterEquation} by expanding all terms in the basis for Hermitian 
matrices, which consists of the identity operator $\mathbb{I}$ and the $SU(N)$ generators $t_i$, with
$i=1,2,\dots,N^2-1$. For this aim, we decompose all operators $\hat{O}$, such as $\hat{\rho}$, $H$ and $\hat{\mathcal{V}}_j$, as
\begin{equation}
\hat{O} = O_0 \mathbb{I} + O_k t_k
\label{basis}
\end{equation}
where the generators $t_i$ satisfy $[t_i,t_j] = i \sum_k f_{ijk}t_k$ and $f_{ijk}$ are the structure constants of $SU(N)$. Then, we get
\begin{equation}
\dot{\rho}_0 = 0\qquad \dot{\rho}_k = \sum_j (M_H)_{kj}\rho_j + \sum_j (M_D)_{jk}\rho_j
\label{MasterEquationComponents}
\end{equation}
being the elements of the matrix $M_H$ given by
\begin{equation}
(M_H)_{kj} = \sum_i h_i f_{ijk}
\end{equation}
where $\rho_k$ and $h_i$ are the components of $\hat \rho$ and $H$, respectively, written on the basis of Equation \eqref{basis}. The matrix $M_D$, which contains all the decoherence parameters, will be called the decoherence matrix. In general, 
it satisfies the following
\begin{enumerate}
\item $M_D=M_D^T$ is a symmetric matrix
\item $-M_D$ is a positive-semidefinite matrix
\item The entries $M_D$ satisfy a set of inequalities inherited from the restrictions on $\hat{\mathcal{V}}_j$
\end{enumerate}
From the second property, it follows that the diagonal elements provide an upper bound to 
the off-diagonal ones. The inequalities mentioned in the third property will only be explicitly written for a few select cases to be discussed
later. We point out that the entries of the decoherence matrix are sometimes assumed to be of the form $(M_D)_{ij} = \gamma_{ij} E^n$, where 
typical values of $n$ are -1,0, and 1. In this paper, we restrict ourselves to analyze the case where $n=0$. 

The component $\rho_0=1/N$ is constant in time and is only relevant when evaluating oscillation probabilities.
The evolution of the $\rho_k$ can then be written in a compact form
\begin{equation}
\dot{\rho} = (M_H+M_D)\rho
\label{MasterEquationMatrix}
\end{equation}
where $\rho$ is an eight-dimensional column vector consisting of the $\rho_k$. The solution to the differential equation is
\begin{equation}
\rho(t) = e^{(M_H+M_D)t}\rho(0)
\label{exponentialEq}
\end{equation}
For a neutrino in a flavour eigenstate $\nu_\alpha$, we can use Eq. \eqref{exponentialEq} to find the evolved vector $\rho_{\nu_\alpha}(t)$.
Finally, the probability $P(\nu_\alpha\to\nu_\beta)\equiv P_{\nu_\alpha\nu_\beta}$  for a neutrino to be detected in the flavour
state $\nu_\beta$ is calculated via inner products
\begin{equation}
P_{\nu_\alpha\nu_\beta} = \frac{1}{N}+\frac{1}{2}(\rho_{\nu_\beta})^T\rho_{\nu_\alpha}(t)
\end{equation}
We emphasize that for neutrinos, which are ultrarelativistic, we have $t=L$ where $L$ is the baseline.

\subsection{Decoherence matrix relation between different quantum bases}
\label{DecoBasisRelations}

When we take into account matter effects in the open neutrino system, it is expected that the decoherence 
matrix in the VMB ($M_D^V$), where the decoherence parameters are defined, has a non-trivial relationship with the one written in the 
MMB ($M_D^M$). The issue of moving to the MMB has been pointed out previously \cite{Oliveira16}. 
However, the description and the consequences of this relationship 
has been overlooked and not properly treated. We will get the connection between both decoherence matrices and explore the validity of 
different forms for the decoherence matrix in the MMB in the context of the neutrino oscillation system. We will also show that some 
decoherence matrices presented in literature are not allowed.   

Starting from the Lindblad master equation and Eq. \eqref{MasterEquationComponents} written in the VMB,
we change to the MMB via the unitary transformation in Eq. \eqref{UnitaryTransformationEq}.
Unitary transformations preserve Hermiticity, so we follow the procedure in section II B and cast the transformed equation
into the form shown in \eqref{MasterEquationComponents}, where the coefficients are now replaced with their corresponding ones 
in the MMB. We also point out that the properties of $M_D$ in section II B are not affected by unitary transformations.

The matrix $M_H$ given in the MMB is the simplest to deal with since the Hamiltonian is diagonal. For the decoherence term, we know that 
$U_T^\dag D[\hat{\rho}] U_T=(M_D^M)_{jk}\rho_j^Mt_k$. On the other hand
\begin{equation}
U_T^\dag D[\hat{\rho}] U_T = (M_D^V)_{jk}\rho_j^V U_T^\dag t_k U_T
\end{equation}
The labels $V,M$ stand for VMB and MMB respectively. 
After a unitary transformation, the Gell-Mann matrices will be a new superposition of the generators $t_k$
\begin{equation}
U_T^\dag t_k U_T = P_{kj}t_j
\end{equation}
where $P_{kj}$ will be an $O(N^2-1)$ matrix. Using this substitution, we find that
\begin{equation}
U_T^\dag D[\hat{\rho}] U_T = \rho_j^M (M_D^M)_{jk} t_k = \rho_j^M P_{ji}(M_D^V)_{il} P_{kl} t_k
\end{equation}
Doing a similar treatment to $\rho_j^M$ and writing it in terms of $\rho_j^V$, we can prove that
the decoherence matrix in the MMB is obtained by performing an orthogonal transformation with the matrix $P$
\begin{equation}
M_D^M = P M_D^V P^T.
\label{NRotationFormula}
\end{equation}
 
From Eq. \eqref{NRotationFormula} a powerful property arises when the elements of the $M_D^M$ are independent of the matter potential 
$A_{CC}$, which implies that they are also independent of energy. In that case,  any value of $A_{CC}$, shall satisfy Eq.
\eqref{NRotationFormula}, in particular $A_{CC}=0$. As we can take $\left.P^T\right|_{A_{CC}=0}=\left.P\right|_{A_{CC}=0}=\mathbb{I}$ we 
have that: 
\begin{equation}
M_D^V=M_D^M, 
\label{DecoMatrixIdentity}
\end{equation}

For instance, the trivial case that fulfills the condition above is when $M_D^V$ is proportional 
to the identity $\mathbb{I}$.

When $M_D^M$= $M_D^V$ the matter+decoherence oscillation probabilities 
can be directly obtained by replacing the standard oscillation angles and masses in the vacuum+decoherence probability formula 
given in \cite{Gago02a} with their effective values in
matter. From now on, this substitution will be referred to as effective matter parameter substitution (EMPS).
We point out that if Eq. \eqref{DecoMatrixIdentity} is not fulfilled, $M_D^M$ is $A_{CC}$ and energy dependent.
 
As we said before, the matrices $M_D^M$ that satisfy Eq.\eqref{DecoMatrixIdentity}, for the case of the potential $A_{CC}$ with constant 
matter density, would also fulfill this equation regardless the value of $A_{CC}$. This observation implies that the aforementioned $M_D^M$ 
matrices are the same even if $A_{CC}(x)$ depends on the position $x$ or the non-constant matter density. In particular, we can conclude that 
these matrices satisfying the condition $M_D^V=M_D^M$ could be used within the framework of the adiabatic or non-adiabatic 
case \cite{Benatti01,Guzzo16}.

We must emphasize that all the results and discussions that will be presented in this paper have been developed within the scheme of a 
potential $A_{CC}$ independent of the position $x$.

\section{TWO GENERATION MIXING}
We first analyze the simple case of two generation mixing to illustrate the effects of rotating the decoherence matrix from the VMB
to the MMB. The Hamiltonian in the VMB $H_V$ and the mixing matrix $U$ are given by
\begin{eqnarray}\nonumber
H_V&=&\Delta \left[\begin{pmatrix} 0&0\\0& 1\end{pmatrix}+U^\dag \begin{pmatrix} A&0\\0&0\end{pmatrix}U\right]\\
U&=&\begin{pmatrix} \cos\theta & \sin\theta\\ -\sin\theta & \cos\theta\end{pmatrix}
\end{eqnarray}
with $\Delta=\Delta m^2/2E$, and $A = V_{CC}/\Delta=A_{CC}/\Delta m^2$. In the MMB, we have the effective value of $\Delta$, given by $\tilde{\Delta}$ and
an effective mixing angle $\tilde{\theta}$ 
\begin{eqnarray}\nonumber
\tilde{\Delta}=\Delta \sqrt{(\cos 2\theta-A)^2+\sin^22\theta}\\
\tan 2\tilde{\theta}=\tan2\theta\left(1-\frac{A}{\cos2\theta}\right)^{-1}
\end{eqnarray}
The matrix $U_M(U_T)$ is obtained directly from $U$ by performing the substitution $\theta\to\tilde{\theta}(\phi)$.
The matrix $P$ that performs the rotation from the VMB to the MMB in the $SU(N)$ basis is
\begin{equation}
P = \begin{pmatrix} \cos 2\phi & 0 & \sin 2\phi\\
0 & 1 & 0\\
-\sin 2\phi & 0 & \cos 2\phi\end{pmatrix}
\end{equation}
Using the notation in \cite{Benatti01}, we can rewrite $P$ using the following correspondences
\begin{equation}
\cos 2\phi = -\frac{\mu}{\sqrt{\mu^2+\nu^2}}\qquad\sin 2\phi = \frac{\nu}{\sqrt{\mu^2+\nu^2}}
\end{equation}
where
\begin{equation}
\mu = \Delta(A\cos 2\theta-1)\qquad\nu = \Delta A\sin 2\theta
\label{DefMuNu}
\end{equation}
Taking a generic decoherence matrix
\begin{equation}
M_D^V =-\begin{pmatrix}
a&b&c\\
b&\alpha&\beta\\
c&\beta&\gamma
\end{pmatrix}
\end{equation}
with $a,\alpha$ and $\gamma$ non-negative, we have the following inequalities that must be satisfied \cite{Benatti01}
\begin{eqnarray}\nonumber
2R\equiv a+\alpha-\gamma\geq 0\qquad & \gamma^2-(a-\alpha)^2 - 4b^2\geq 0\\\nonumber
2S\equiv a+\gamma-\alpha\geq 0\qquad & \alpha^2-(a-\gamma)^2 - 4c^2\geq 0\\\nonumber
2T\equiv \alpha+\gamma-a\geq 0\qquad & a^2-(\alpha-\gamma)^2 -4\beta^2\geq 0
\end{eqnarray}\begin{eqnarray}
RST-2bc\beta-R\beta^2-Sc^2 -Tb^2\geq 0
\label{Decoinequalities}
\end{eqnarray}
The corresponding matrix in the MMB is
\begin{equation}
M_D^M=-\begin{pmatrix}
\tilde{a}&\tilde{b}&\tilde{c}\\
\tilde{b}&\alpha&\tilde{\beta}\\
\tilde{c}&\tilde{\beta}&\tilde{\gamma}
\end{pmatrix}
\end{equation}
Given the form of $P$, we see that $\alpha$ is naturally not affected, while the effective decoherence parameters are
\begin{eqnarray}\nonumber
\tilde{a}&=&\frac{a+\gamma}{2}+\frac{a-\gamma}{2}\cos 4\phi-c\sin4\phi\\\nonumber
\tilde{b}&=&b\cos2\phi-\beta\sin2\phi\\\nonumber
\tilde{c}&=&c\cos4\phi +\frac{a-\gamma}{2}\sin4\phi\\\nonumber
\tilde{\beta}&=&\beta\cos2\phi+b\sin2\phi\\
\tilde{\gamma}&=&\frac{a+\gamma}{2}-\frac{a-\gamma}{2}\cos 4\phi+c\sin4\phi
\label{2GenEffDecoParams}
\end{eqnarray}
From this equation, we see that a diagonal decoherence matrix in the VMB, which is $b=c=\beta=0$, can have off-diagonal entries in the MMB. 
Also, since the matrix $P$ depends on $\tilde{\theta}$, the decoherence matrix in the MMB is inherently dependent on $E$ and $A$. 
It is interesting to note that some models provide a power law energy dependence to the decoherence parameters, which will not hold
in the MMB.

We continue by analyzing some features of $2\nu$ oscillations in the presence of decoherence. We will also present a few numerical examples,
which will use the standard oscillation parameters summarized in Table \ref{Table1}. Decoherence parameters are assigned on a case-by-case
basis.
\begin{table}\begin{center}\begin{tabular}{c|c}
Parameter & Value \\\hline
$\theta$ & $9^\circ$\\
$\Delta m^2$ & $2.3\times 10^{-3}$ eV$^2$\\
Matter Density $\rho$ & $2.97$ g cm$^{-3}$\\
Baseline $L$ & 1300km
\end{tabular}
\caption{Standard oscillation parameters used for $2\nu$ oscillation examples.}
\label{Table1}
\end{center}\end{table}

\subsection{Conditions for effective matter parameter substitution (EMPS)}
From Eq. \eqref{2GenEffDecoParams}, we see that 
the only decoherence matrix that satisfies Eq. \eqref{DecoMatrixIdentity}
for the two neutrino generations case is:
\begin{equation}
M_D^V = -\text{Diag}(\gamma,\alpha,\gamma)=M_D^M
\end{equation}
constrained by the condition $\alpha\leq 2\gamma$. 

As we have already mentioned, when Eq. \eqref{DecoMatrixIdentity} is fulfilled the matter oscillation+decoherence probabilities can be easily obtained 
through the EMPS (i.e. $\Delta\to\tilde{\Delta},\theta\to\tilde{\theta}$). 

\begin{figure}
\includegraphics[width=0.5\textwidth]{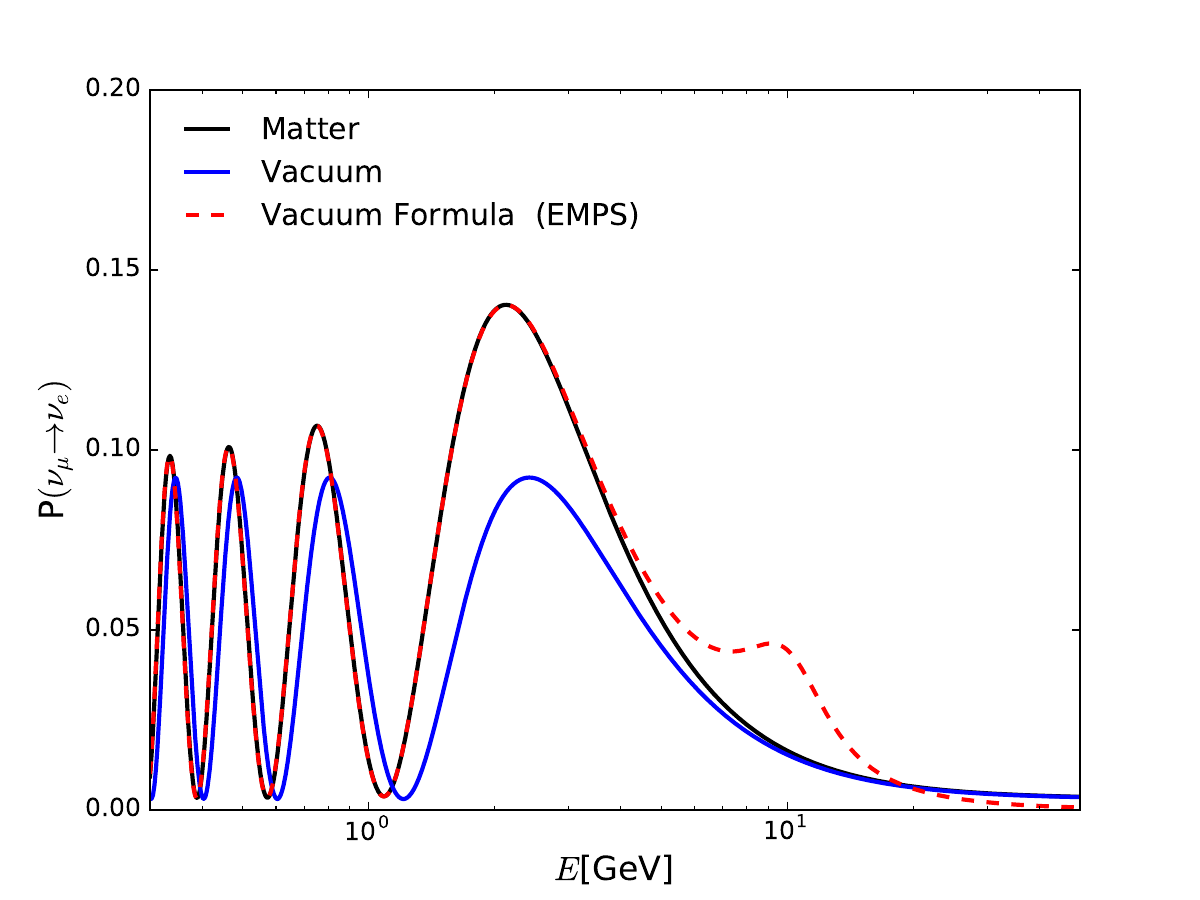}
\caption{Example of misuse of the EMPS on the transition probability for $M_D^V=-\text{Diag}(a,a,0)$ with $a=1\times 10^{-23}$ GeV and with the other parameters fixed according
to the values given at Table ~\ref{Table1}. The vacuum and matter oscillation probability are also included and were calculated numerically.}
\label{2GenComp}
\end{figure}

In order to exemplify what happens when the aforementioned replacement is misused, we take the following decoherence matrix, given in 
\cite{Oliveira16}: 
\begin{equation}
M_D^V= -\;\;\text{Diag}(a,a,0)
\end{equation}
This matrix, written in the MMB is:
\begin{equation}
M_D^M=-\begin{pmatrix}
a\cos^22\phi & 0 & \frac{a}{2}\sin 4\phi\\
0 & a & 0\\
\frac{a}{2}\sin 4\phi & 0 & a\sin^22\phi
\end{pmatrix}
\end{equation}
Then, the identity given in Eq. \eqref{DecoMatrixIdentity}, that allows us to make the EMPS, breaks down. This is ilustrated through 
three curves presented in  Figure \ref{2GenComp}: one is the vacuum+decoherence oscillation probability; the other is the exact  matter + 
decoherence oscillation probability, obtained numerically; 
the last one is doing the EMPS, this is the substitution $\Delta\to\tilde{\Delta}$ and $\theta\to\tilde{\theta}$ into the vacuum formula
$P_{\nu_\mu\nu_e} = \sin^22\theta\left(1-e^{-a t}\cos\Delta t\right)/2 $.  
Comparing these curves, we see how the EMPS fails to describe
the decoherence phenomenon, producing a fake peak in the 
probability around the energy region where $\tilde{\theta}=\pi/4$ that allows maximal neutrino mixing. 
 At lower energies we observe that the exact matter+decoherence formula and the direct susbtitution formula coincide very well. 
This is explained because at this part of the energy spectrum we have that  $\tilde{a}\rightarrow {a}$, $\tilde{\gamma}=\tilde{b}=
\tilde{\beta}=\tilde{c}=0$, making the direct substitution valid for these energies.    

\subsection{Relevance of off-diagonal decoherence parameters}
As we have pointed out in section \ref{DensityMatrixFormalism}, the decoherence matrix parameters are bound by the diagonal elements. 
Based on this, it is tempting to say that off-diagonal parameters can be neglected. In light of Eq. \eqref{2GenEffDecoParams},
we see that in the MMB the diagonal entries of the decoherence matrix also receive contributions from $c$.

We revisit this point to view the impact of off-diagonal parameters on oscillation probabilities, by considering the matrix proposed in \cite{Benatti01}
\begin{equation}
M_D^V= -\begin{pmatrix}
\gamma&b&0\\
b&\gamma&\beta\\
0&\beta&\gamma
\end{pmatrix}
\end{equation}
with $b^2+\beta^2\leq \gamma^2/4$, from which the following transition probability is obtained
\begin{equation}
P_{\nu_\mu\nu_e} = \frac{1}{2}\left(1-e^{-\gamma t}\right)+\left[\frac{\bar{\nu}^2-\bar{\beta}^2}{\Omega^2}\right]e^{-\gamma t}
\sin^2 \left(\frac{\Omega t}{2}\right)
\end{equation}
where we defined
\begin{displaymath}
\bar{\nu} = \Delta\sin 2\theta\qquad\qquad \bar{\beta}=\beta\cos2\theta+b\sin2\theta
\end{displaymath}\begin{equation}
\Omega = \left(\mu^2+\nu^2-b^2-\beta^2\right)^{1/2}
\end{equation}
The interesting feature is that $b,\beta$ affects the oscillation frequency. A simple extension where $c\neq 0$ is
\begin{equation}
M_D^V= -\begin{pmatrix}
\gamma&b&c\\
b&\gamma&\beta\\
c&\beta&\gamma
\end{pmatrix}
\label{MatrixOffDiag}
\end{equation}
with the following constrains
\begin{eqnarray}\nonumber
b^2\leq\gamma^2/4 \qquad\beta^2\leq\gamma^2/4\qquad c^2\leq \gamma^2/4\\
\frac{\gamma}{2}\left(\frac{\gamma^2}{4}-(b^2+c^2+\beta^2)\right)-2bc\beta\geq 0
\end{eqnarray} 
\begin{figure}
\includegraphics[width=0.5\textwidth]{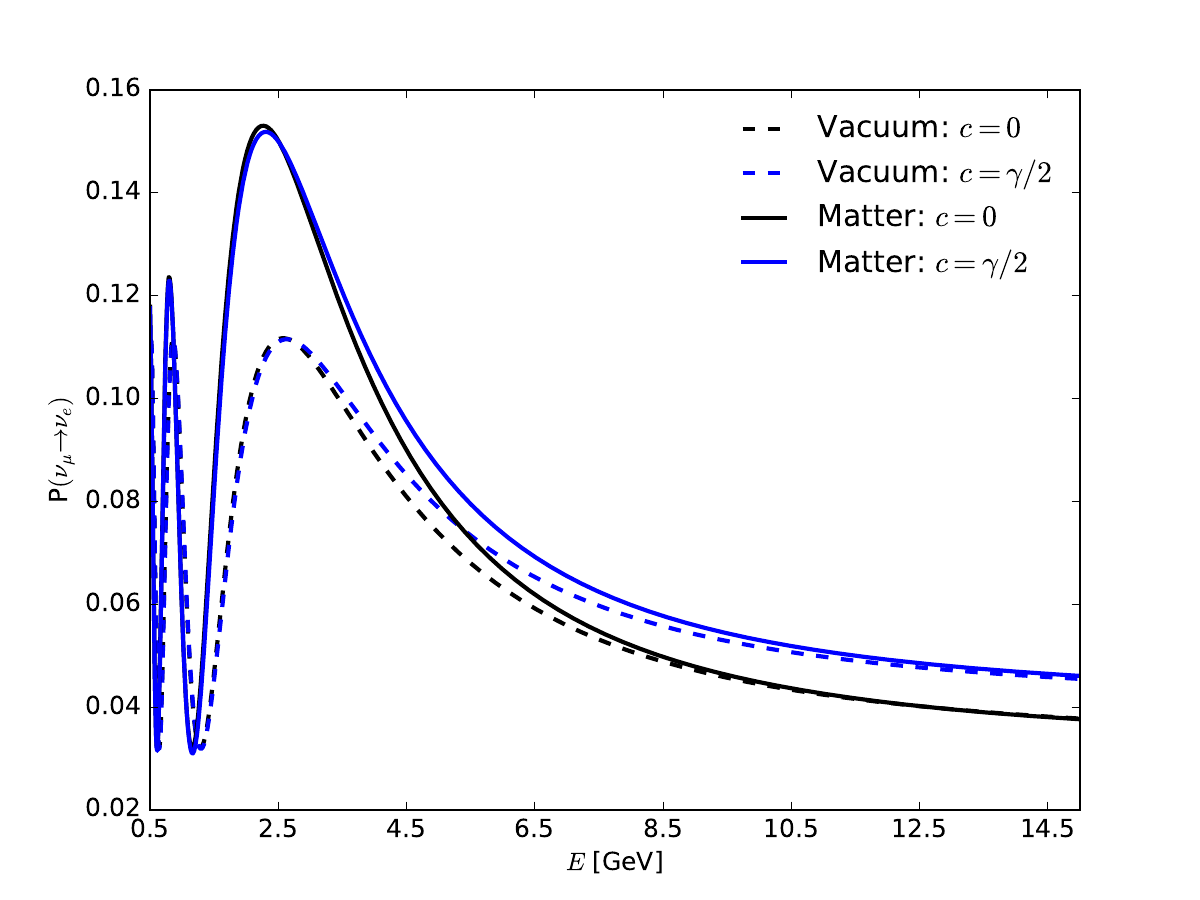}
\caption{Oscillation probabilities for the decoherence matrix given in \eqref{MatrixOffDiag}. All the probabilities presented in the plot were numerically found.}
\label{2Gen_DiagCompFig}
\end{figure}

For this scenario, there is no simple form for the matrix exponential (Eq. \eqref{exponentialEq}), even if $b=\beta=0$ or $A=0$. 
An alternative is to assume that the decoherence parameters are small $\gamma\ll \Delta,\gamma t\ll 1$. Expanding to first order in the decoherence parameters
\begin{eqnarray}\nonumber
P_{\nu_e\nu_\mu}&=&\sin^22\tilde{\theta}\left[\sin^2\frac{\Omega t}{2}+\left(\gamma+\frac{c\mu\nu}{\Omega^2}\right)t\cos\Delta t
+\frac{c\mu\nu}{\Delta\Omega^2}\right]\\\nonumber
& & +\frac{1}{2}\cos^22\tilde{\theta}\left(\gamma-\frac{c\mu\nu}{\Omega^2}\right)t\\
& &+\frac{c}{2\Delta}\sin4\tilde{\theta}\frac{\mu^2-\nu^2}{\Omega^2}\sin\Delta t
\end{eqnarray}
where $\mu,\nu$ are defined in \eqref{DefMuNu} and $\Omega = (\mu^2+\nu^2)^{1/2}$.
Note that the parameters $b,\beta$ do not contribute to the oscillation probability at first order. This feature is independent of the 
particular form of $M_D^V$. The vacuum limit is recovered by setting $\mu\to-\Delta,\nu\to0,\Omega\to\Delta$ and $\tilde{\theta}
\to\theta$
\begin{eqnarray}\nonumber
P_{\nu_e\nu_\mu}^{A=0}&=&\frac{1}{2}\gamma t+\sin^22\theta(1-\gamma t)\sin^2\frac{\Delta t}{2}\\
& & +\frac{c}{2\Delta}\sin 4\theta\sin\Delta t
\end{eqnarray}
We appreciate that the off-diagonal parameter $c$ has a contribution comparable to the diagonal element $\gamma$ even in the vacuum. 
In Figure~\ref{2Gen_DiagCompFig}, we present the probability when $b=\beta=0$ and $\gamma=10^{-23}$ 
GeV, which is appropriate for the
DUNE baseline. The values chosen for $c$ are zero and $\gamma/2$ and we display the results after propagation in vacuum and in matter. 
In both cases, it is clear that the contribution from $c$ cannot be neglected. 
The discrepancy between $c=0,c=\gamma/2$ starts from approximately 4.0 GeV and increases with energy. 

\subsection{Conservation of energy}
In literature, the possibility of choosing a decoherence matrix such that energy is conserved has also been suggested. Conservation of
energy implies that Tr$(\dot{\rho}H)=0$ which requires $\dot{\rho}_3^M=0$ if we work in the MMB. This conditions reads
\begin{equation}
-\tilde{c}\rho_1^M-\tilde{\beta}\rho_2^M-\tilde{\gamma}\rho_3^M=0
\end{equation}

For this to be valid at all $t$ requires $\tilde{c}=\tilde{\beta}=\tilde{\gamma}=0$. We also need these three parameters to be $E$-independent,
meaning that $a=b=c=\beta=\gamma=0$ and $\alpha$ arbitrary (see Eq. \eqref{2GenEffDecoParams}). The decoherence matrix
must also satisfy the inequalities $2R \geq 0$ and  $2S \geq 0$, given in Eq. \eqref{Decoinequalities}, which is only achieved if $\alpha=0$ (the remaining inequalities are automatically satisfied). 
We therefore conclude that there is no decoherence phenomenon that allows for energy conservation in matter at all energies.

\section{THREE GENERATION MIXING}
\label{3NuDecoherence}
\subsection{Perturbative approach}
We now write the decoherence matrix in the VMB as $M_D^V = \Gamma M_\Gamma^V$, where $\Gamma = $max$|(M_D)_{ii}|$. In order to develop our pertubative approach we treat $\Gamma L$ as small parameter, with $\Gamma L\sim$ 0.1 being the 
upper limit for the validity of our probability formula. For a 
DUNE baseline of 1300 km this corresponds to $\Gamma \sim 10^{-23}$GeV. Likewise, the leading term in $M_H$ is of order $\mathcal{O}(\Delta_{31})$ and, for an energy $E=10$ GeV, $\Gamma/\Delta_{31}\sim 0.08$.

The matrix $U_T$ uses the effective oscillation parameters, which are expanded in power series of $\alpha = \Delta m^2_{21}/\Delta m^2_{31}$
and $\theta_{13}$ (see Eqs.~\ref{EffectiveAngleFormula} and~\ref{EffectiveMassFormula}). The matrix $M_D^M=PM_D^VP^T$ will admit a 
similar expansion
\begin{equation}
M_D^M = PM_D^VP^T=\Gamma M_\Gamma + \Gamma\alpha M_{\Gamma\alpha} + \Gamma\theta_{13}M_{\Gamma\theta} + \dots
\label{DecoMatrixDecomposition}
\end{equation}
The standard oscillation contribution can be solved exactly in the MMB, so we re-write equation \eqref{MasterEquationMatrix} as
\begin{equation}
\dot{\tilde{\rho}}(t) = (\Gamma \tilde{M}_\Gamma + \Gamma\alpha \tilde{M}_{\Gamma\alpha} + \Gamma\theta_{13}\tilde{M}_{\Gamma\theta}+
\dots)\tilde{\rho}(t)
\label{DiffEq3Gen}
\end{equation}
where $\tilde{\rho}(t)=e^{-M_Ht}\rho^M(t)$ and the matrices $\tilde{M}$ are related to the previous $M$ via $\tilde{M} = e^{-M_Ht}Me^{M_Ht}$.
This equation is then solved perturbatively by expanding the solution $\tilde{\rho}$ as a power series,
\begin{eqnarray}\nonumber
\tilde{\rho} &=& \rho^{(0)}+\alpha\rho^{(\alpha)}+\theta_{13}\rho^{(\theta)}+\Gamma\rho^{(\Gamma)}\\\nonumber
& & +\alpha\theta_{13}\rho^{(\alpha\theta)}+\theta_{13}^2\rho^{(\theta^2 )}+\alpha^2\rho^{(\alpha^2)}\\
& & +\Gamma\theta_{13}\rho^{(\Gamma\theta)}+\Gamma\alpha\rho^{(\Gamma\alpha)}+\Gamma^2\rho^{(\Gamma^2)}+\dots
\label{Solution3Gen}
\end{eqnarray}
yielding a sequence of first order differential equations by substituting \eqref{Solution3Gen} into \eqref{DiffEq3Gen}
and grouping terms of equal powers. The solutions can be found with software such as Mathematica and will only
be shown for a particular decoherence matrix in Sec \ref{3GenOurModel}. 

We must note that all $\Gamma$-independent terms ($\rho^{(0)},\rho^{(\theta)}$,etc.)
in the solution of \eqref{Solution3Gen} do not evolve in time and will contribute to the standard neutrino oscillation probability in matter. 
Although we do not write these terms explicitly in the probability formula, since the standard oscillation contribution is calculated
numerically, these $\Gamma$-independent terms are still required to calculate terms of higher order that involve $\Gamma$. 
For example, we need to find $\rho^{(\alpha)}$ to calculate $\rho^{(\Gamma\alpha)}$.

The initial condition $\tilde{\rho}(0)=\rho(0)=\rho_{\nu_\beta}$ depends on $U_M$ only and is $\Gamma$-independent.
Any term proportional to $\Gamma$ ($\rho^{(\Gamma)},\rho^{(\Gamma\alpha)}$, etc.) will therefore vanish at $t=0$. 
After calculating $\tilde{\rho}$, oscillation probabilities are found using 
\begin{equation}
P_{\nu_\beta\nu_{\beta'}} = \frac{1}{3}+\frac{1}{2}(\rho_{\nu_{\beta'}}^M)^Te^{M_Ht}\tilde{\rho}_{\nu_\beta}(t)
\end{equation}
Given our method, the probability is also a power series in $\alpha,\theta_{13}$ and the elements in the decoherence matrix.
 
\subsection{Decoherence matrix}
Starting from a diagonal matrix $M_D^V$, with all its elements different, and demanding $M_D^V=M_D^M$ in Eqs.~\eqref{off1stord} and~\eqref{off2ndord} given in Appendix~\ref{MMBrotations_analytical} we find that for $\delta=0,\pi$:
\begin{equation}
M_D^V=M_D^M =-\text{Diag}(\Gamma_1,\Gamma_2,\Gamma_1,\Gamma_1,\Gamma_2,\Gamma_1,\Gamma_2,\Gamma_1),
\label{diagvalid0}
\end{equation}  
and for $\delta=\pi/2,3\pi/2$:
\begin{equation}
M_D^V =M_D^M = -\text{Diag}(\Gamma_1,\Gamma_2,\Gamma_1,\Gamma_2,\Gamma_1,\Gamma_2,\Gamma_1,\Gamma_1)
\label{diagvalidpi2}
\end{equation}
subject to the condition $\Gamma_1/3\leq \Gamma_2\leq 5\Gamma_1/3$.

We have verified numerically that the diagonal cases above are the only ones that remain unchanged after being rotated into the MMB. 
Besides these cases we have the trivial one when $M_D^V$ is proportional to the $\mathbb{I}$. As we have already mentioned in 
Sec \ref{DecoBasisRelations}, only for these diagonal $M_D^M$ matrices we can apply the EMPS to obtain the matter+decoherence oscillation 
probabilities. 

We notice that the latter procedure has been misleadingly applied in recent papers. For instance, the $M_D^M$ given in \cite{Oliveira16}:
\begin{equation}
M_D^M=-\text{Diag}(\Gamma_1,\Gamma_1,0,\Gamma_2,\Gamma_2,\Gamma_3,\Gamma_3,0),
\label{diagvalid_Oliveira}
\end{equation}
or 
\begin{equation}
M_D^M=-\text{Diag}(0,0,0,\Gamma,\Gamma,\Gamma,\Gamma,0),
\label{diagvalid_PRL}
\end{equation}
used in \cite{Coelho17} do not depend on $A$ or $E$ and must follow the relation
$M_D^V=M_D^M$. The matrix changes form when rotated back to the VMB and the resulting
$M_D^V$ is $A$-dependent.
The latter does not make sense since decoherence, which is actually defined in the VMB, is an effect independent of the matter potential. 
Therefore the matrices  $M_D^M$ given in Eqs. \eqref{diagvalid_Oliveira} and \eqref{diagvalid_PRL} can not be derived from a realistic 
decoherence scenario. On the other hand, if we consider the matrices  in Eqs. \eqref{diagvalid_Oliveira} and \eqref{diagvalid_PRL} 
as $M_D^V$, decoherence matrices  defined in the VMB,  they certainly meet all the physical conditions required to be a decoherence matrix in 
vacuum. However, they are not suitable for the application of the EMPS.

For simplicity, we will assume for our calculations the decoherence matrix in the VMB as diagonal. We assume one of the simpler forms
\begin{equation}
M_D^V = -\text{Diag}(\Gamma_2,\Gamma_2,0,\Gamma_4,\Gamma_4,\Gamma_4,\Gamma_4,0)
\label{Model1Eq}
\end{equation}
which is subject to the condition $\Gamma_2\leq 4\Gamma_4$. 
\subsection{Matter+decoherence oscillation probabilities}
\label{3GenOurModel}

\begin{figure*}[t]
\centerline{
\includegraphics[width=0.5\textwidth]{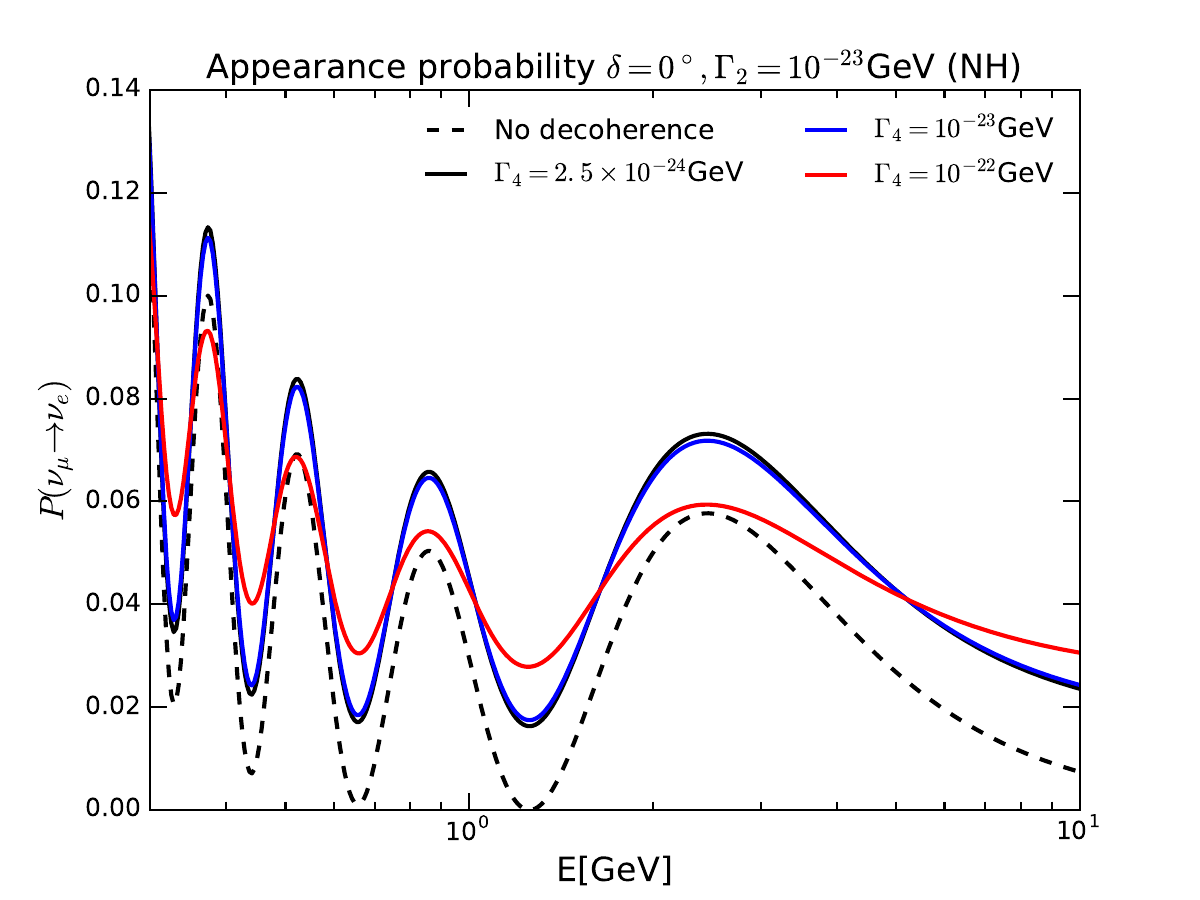}
\includegraphics[width=0.5\textwidth]{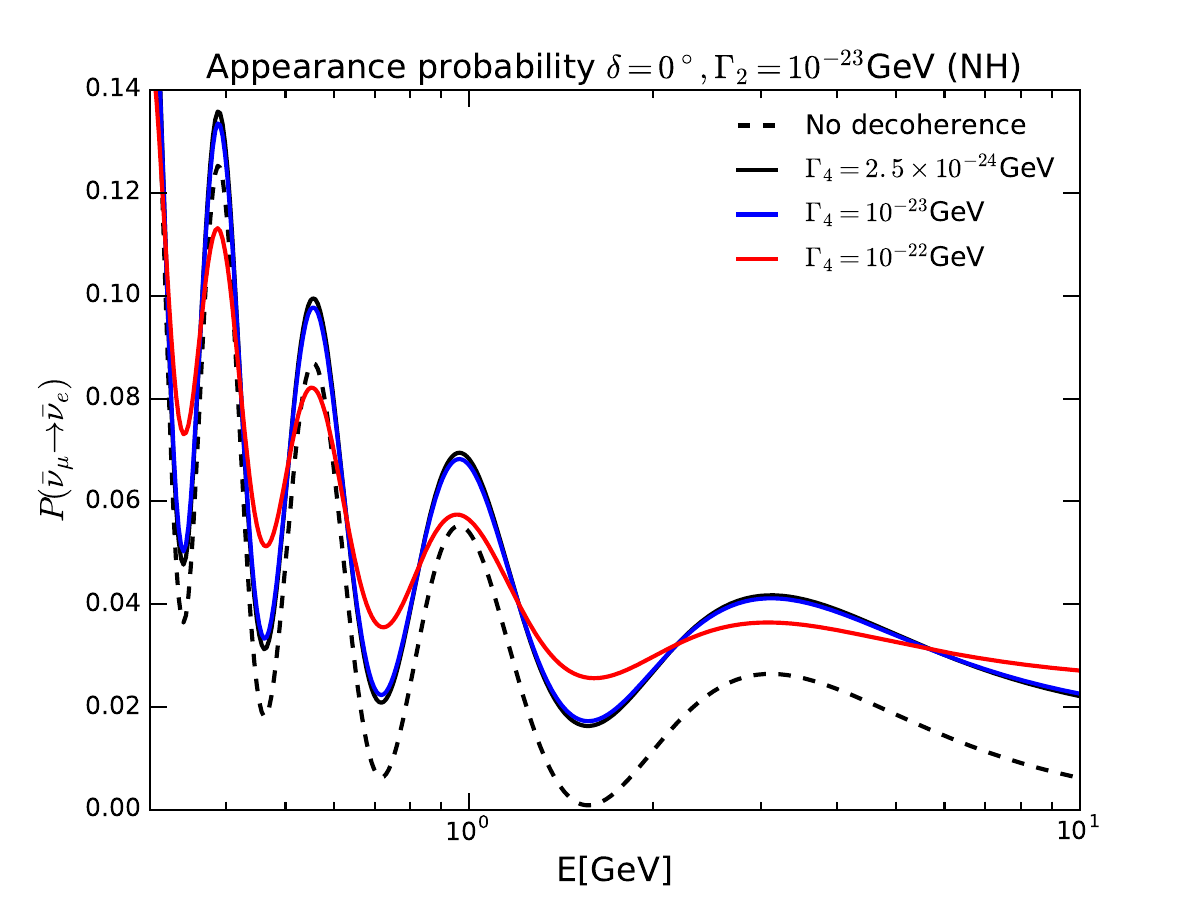}
}
\caption{Effects of decoherence on appearance probabilities assuming DUNE's baseline, $\delta=0^\circ$ and normal hierarchy
for neutrino (left) and antineutrino (right) channels.}
\label{3NuProbPlots}
\end{figure*}

We set $t=L$ and, in the context of DUNE, $L=1300$ km is going to be the baseline of the experiment.
As a first scenario, we will assume there is no decoherence and define $\Delta=\Delta_{31}L/2$. Defining $\bar{\Gamma}_i=\Gamma_iL$,
we obtain the neutrino oscillation probabilities, where the appearance probability is given by
\begin{eqnarray}\nonumber
P_{\nu_\mu\nu_e}&=&P_{\nu_\mu\nu_e}^{(0)}+\frac{1}{2}\bar{\Gamma}_2\cos^2\theta_{23}\sin^22\theta_{12}\\\nonumber
& &-\frac{1}{4}\cos^2\theta_{23}\bar{\Gamma}_2^2\left[\sin^42\theta_{12}+\sin^24\theta_{12}\frac{\sin^2(A\Delta)}{4A^2\Delta^2}\right]\\\nonumber
& &+\frac{\bar{\Gamma}_2\theta_{13}\sin2\theta_{23}\sin4\theta_{12}}{4(1-A)A\Delta}\left[\sin(A\Delta)\cos(\delta+A\Delta)\right.\\\nonumber
& &\left.\qquad-A^2\sin\Delta\cos(\delta+\Delta)\right]\\\nonumber
& &-\frac{\alpha\bar{\Gamma}_2}{2A^2\Delta}\cos2\theta_{12}\cos^2\theta_{23}\sin^22\theta_{12}\\
& &\qquad\times \left(\sin2A\Delta-2A\Delta\right)
\label{AppProb_Analytic}
\end{eqnarray}
and the survival probability is
\begin{eqnarray}\nonumber
P_{\nu_\mu\nu_\mu}&=&P_{\nu_\mu\nu_\mu}^{(0)}-\frac{1}{2}\sin^22\theta_{23}\left(\bar{\Gamma}_4-\frac{1}{2}\bar{\Gamma}_4^2\right)\cos(2\Delta)\\\nonumber
& &+\frac{1}{4}\cos^4\theta_{23}\bar{\Gamma}_2^2\left[\sin^42\theta_{12}+\sin^24\theta_{12}\frac{\sin^2(A\Delta)}{4A^2\Delta^2}\right]\\\nonumber
& &+\frac{\bar{\Gamma}_2\alpha\cos2\theta_{12}\cos^4\theta_{23}\sin^22\theta_{12}}{2A^2\Delta}(\sin2A\Delta-2A\Delta)\\\nonumber
& &-\frac{\bar{\Gamma}_2\theta_{13}\sin2\theta_{23}\cos^2\theta_{23}\sin4\theta_{12}\cos\delta}{4(1-A)\Delta}\\
& &\qquad\times(\sin2A\Delta-A^2\sin2\Delta)
\label{DisProb_Analytic}
\end{eqnarray}
The probabilities $P_{\nu_\mu\nu_e}^{(0)}$, $P_{\nu_\mu\nu_e}^{(0)}$ are, respectively, the appearance and survival probabilities in the 
absence of decoherence ($\Gamma_i=0$). These can be calculated numerically by any standard neutrino oscillation package. The antineutrino
channels are found via the replacement $\delta\to -\delta$ and $A\to -A$. We have also 
compared this result to numerical simulations, with an error lower than 5\%  in the energy range $0.3\leq E/$GeV$\leq10.0$ when $\Gamma\leq
10^{-23}$ GeV for all channels, assuming the central values of the standard oscillation parameters. We therefore
use these formulas to describe new features arising from decoherence. 

In this case, we see that $\Gamma_4$ does not contribute to the survival probability at the lowest order.
Note that both the appearance and survival probabilities exhibit $A$ and $E$-independent terms proportional to 
$\Gamma_2$ or $\Gamma_2^2$. We will call these effects a shift. A second term proportional to $\Gamma_2^2$ is also present, which depends on
$\sin(A\Delta)$, and becomes a matter-dependent term that vanishes for vacuum oscillations. 

One of the more interesting features is CP violation induced by decoherence. Looking at Eq. \eqref{AppProb_Analytic}, the CP violating
terms are proportional to $\Gamma_2\alpha,\Gamma_2\theta_{13}$, in addition to the standard CP violation arising from 
$P_{\nu_\mu\nu_e}^{(0)}$. In the limit $A\to 0$, decoherence induced CP violating terms vanish at second order and the effect 
becomes subdominant. 

\subsection{Results}
To present our results, we calculate the $3\nu$ neutrino oscillation probabilities, including decoherence effects, using the nuSQuIDS package 
\cite{Arguelles14}. We compare the case of no decoherence with the decoherence bi-parameter model given in Eq. \eqref{Model1Eq}. We set 
$\Gamma_2=10^{-23}$ GeV and $\Gamma_4$ to the following values: $2.5 \times 10^{-24}, 10^{-23}$ and $10^{-22}$ GeV. These values of $\Gamma_4$
are chosen in order to satisfy $\Gamma_2 \leq 4\Gamma_4$, where $\Gamma_4= 2.5 \times 10^{-24}$ GeV is the limit value. Standard neutrino 
oscillation parameters are fixed to the central values in \cite{Acciarri15} (see Table \ref{Table2}) with the exception of $\delta$. Normal 
hierarchy (NH) is assumed throughout. 

\begin{table}\begin{center}\begin{tabular}{c|c|c}
Parameter & Value \\\hline
$\theta_{12}$ & 0.5843\\
$\theta_{23}$(NH) & 0.738 \\
$\theta_{13}$ & 0.148 \\
$\Delta m^2_{21}$ & $7.5\times 10^{-5}$ eV$^2$\\
$\Delta m^2_{31}$ (NH) & $2.457\times 10^{-3}$ eV$^2$ \\
Matter Density $\rho$ & $2.97$ g cm$^{-3}$\\
Baseline $L$ & 1300km
\end{tabular}
\caption{Standard neutrino oscillation parameters obtained from global fits \cite{Acciarri15} and DUNE baseline parameters.}
\label{Table2}
\end{center}\end{table}

The appearance probabilities with the aforementioned set of values are shown in Figure \ref{3NuProbPlots}, for both neutrinos and 
antineutrinos.
As we can see the appearance probability has no significant change in shape after introducing decoherence. We do observe that the
probabilities are higher in the presence of decoherence at all energies, whether it is the neutrino or antineutrino channel. 
The term responsible for this shift is the energy independent term proportional to $\Gamma_2$ which dominates the decoherence contribution 
appearing in our \eqref{AppProb_Analytic}. It is important to recall  that this formula is valid up to values $\Gamma \sim 10^{-23}$ GeV.  
We also note that for values of $\Gamma_4$ between $2.5 \times 10^{-24}$ and $10^{-23}$ GeV, this parameter has a negligible influence in the 
probabilities. Our analytical formula \eqref{AppProb_Analytic} reflects this since it has no decoherence contribution involving $\Gamma_4$. 
On the other hand, when $\Gamma_4 $ is raised to $10^{-22}$ GeV, past the validity of our formula, the damping effect due to decoherence 
beigns to take control in the oscillation probabilities, diminishing its amplitudes.
\begin{figure*}
\vspace{-4pt}
\centerline{
\includegraphics[width=0.5\textwidth]{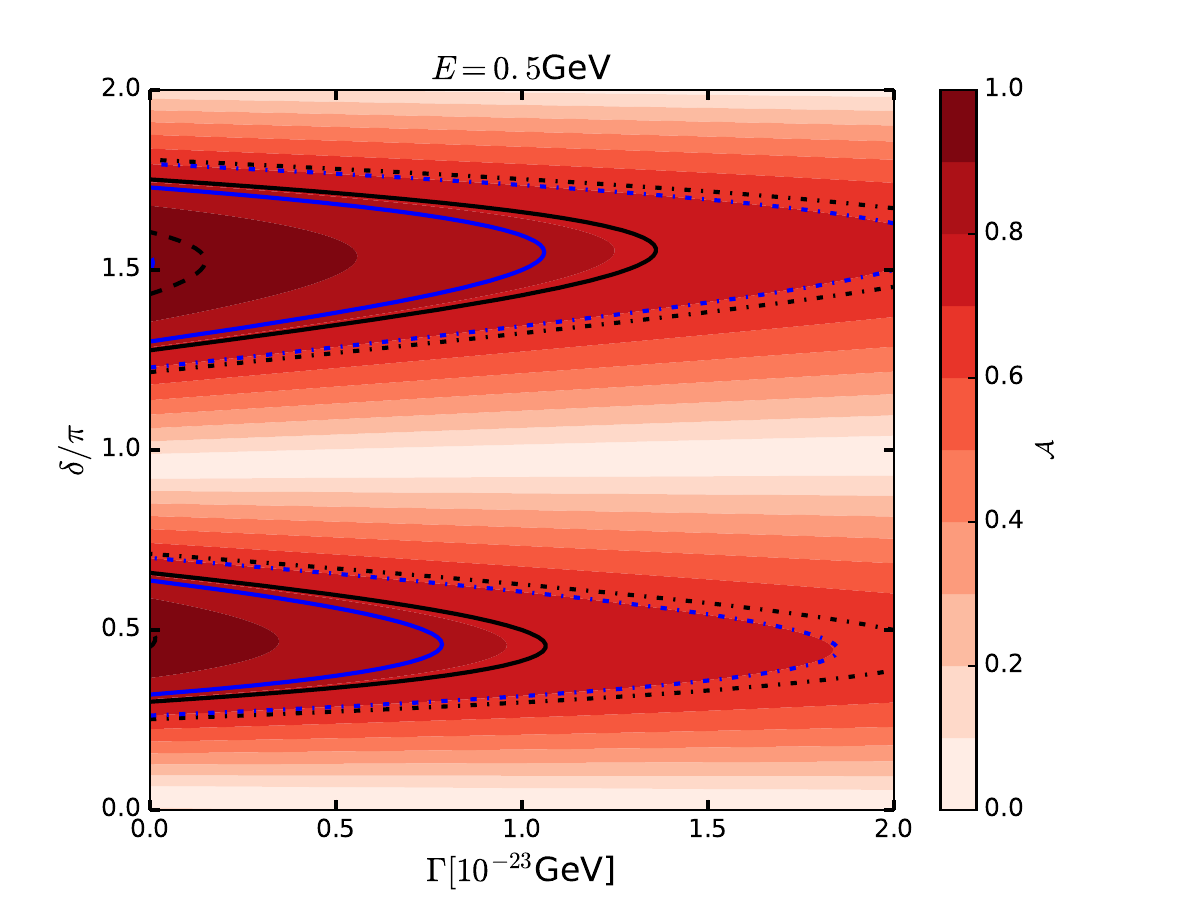}
\includegraphics[width=0.5\textwidth]{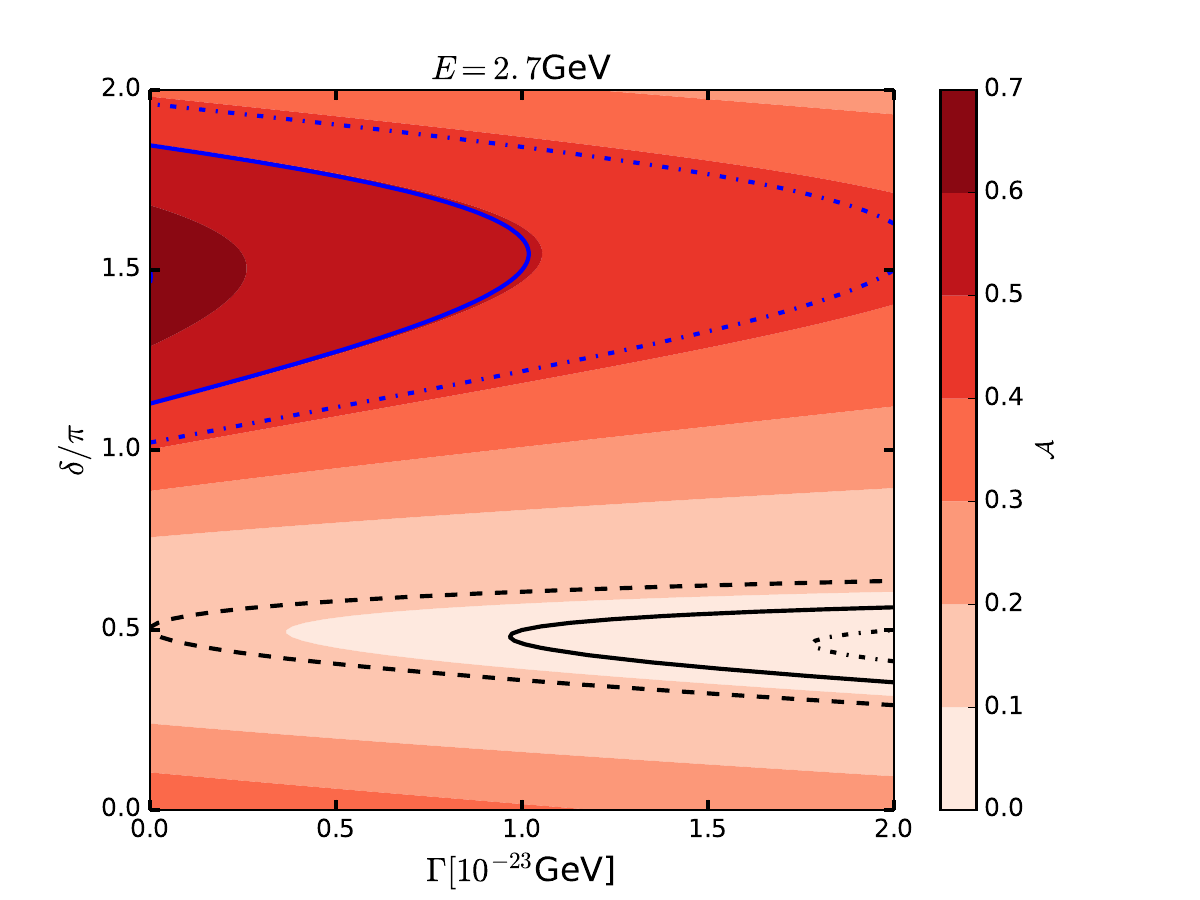}
}
\centerline{
\includegraphics[width=0.5\textwidth]{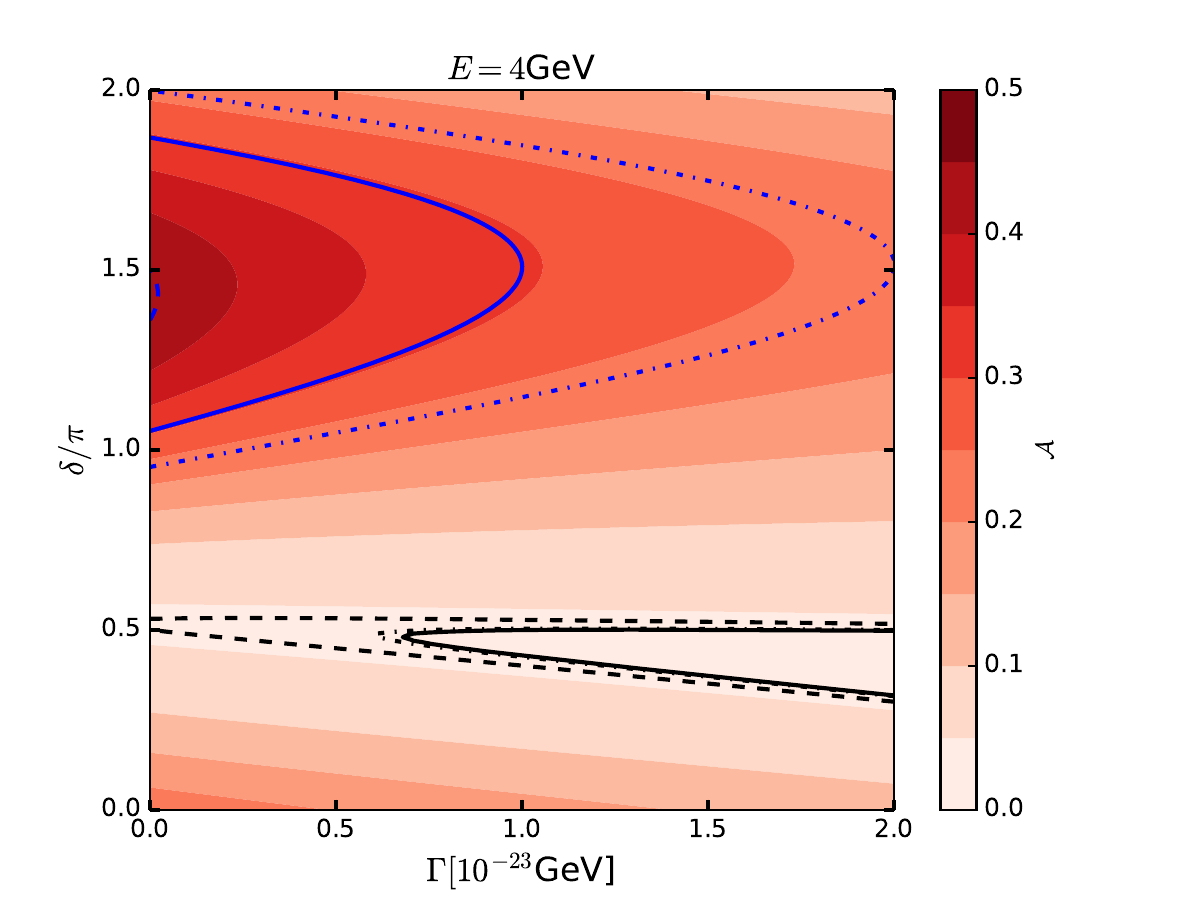}
\includegraphics[width=0.5\textwidth]{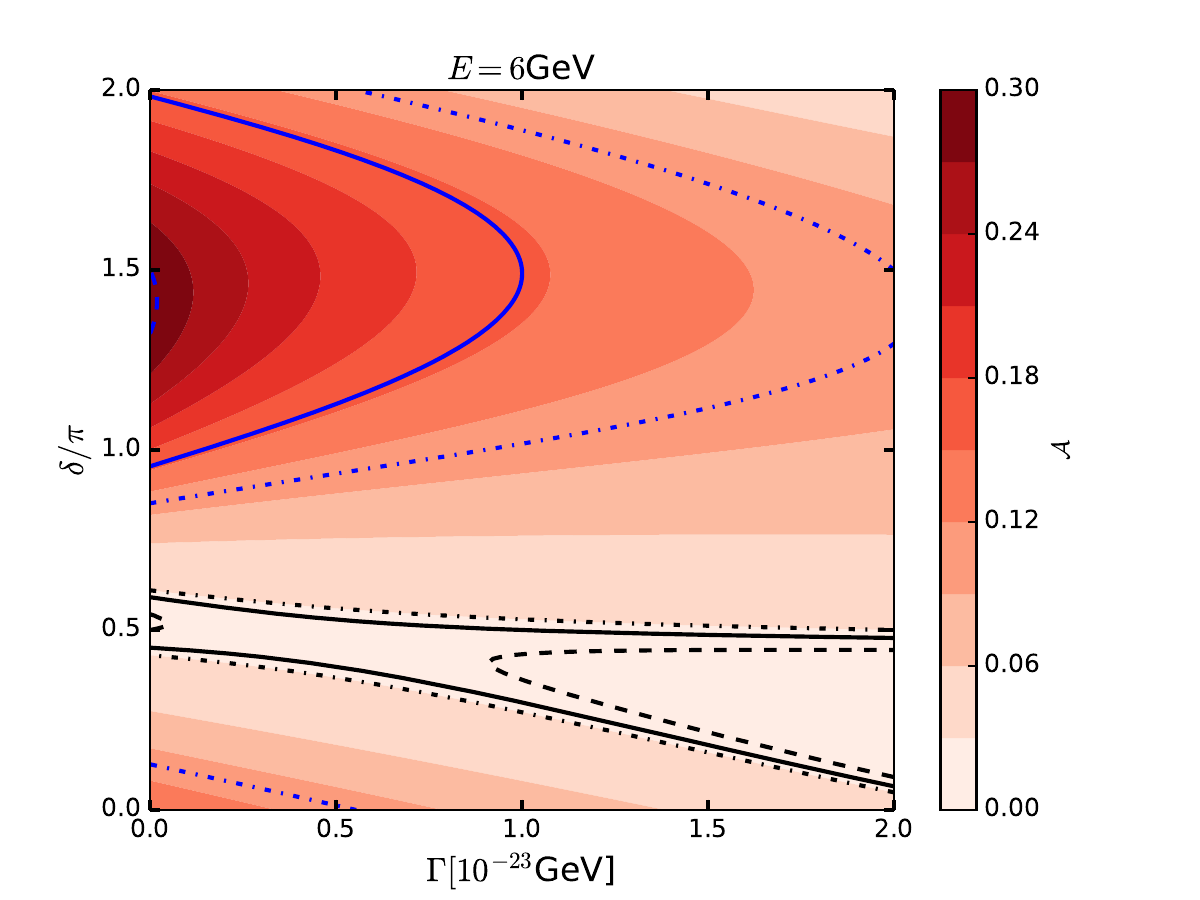}
}

\caption{Effects of decoherence on CP assymetry, assuming DUNE's baseline and normal hierarchy for various energies. 
The blue (black) lines correspond to $\delta=3\pi/2(\pi/2)$ contours at select values of the decoherence 
parameter $\Gamma$, using solid (dot-dashed) lines for $\Gamma=10^{-23}\text{GeV} (2\times10^{-23}\text{GeV})$. The dashed line is the 
$(\pi/2, 0)$ contour.}
\label{3NuCPPlots}
\end{figure*}

Additionally, using our numerical approach, we have tested the results given in \cite{Oliveira16}. We do not reproduce the different peaks 
exhibited in the plots given in this reference, and its probabilities do not have the energy independent shift predicted by our numerical and 
analytical result (see Figure 4 in \cite{Oliveira16}). Similarly to the two generation case, in \cite{Oliveira16} the effects of rotating
the decoherence matrix in the three generation framework were not been properly treated. The latter is the root of the appearence of these 
strange peaks.       

After assessing the decoherence effects on the probabilities, we look into   
its effects on CP violation. For this study we use the CP asymmetry defined as
\begin{equation}
\mathcal{A}=\frac{|P(\nu_\mu\to\nu_e)-P(\bar{\nu}_\mu\to\bar{\nu}_e)|}{P(\nu_\mu\to\nu_e)+P(\bar{\nu}_\mu\to\bar{\nu}_e)}
\end{equation}

In Figure \ref{3NuCPPlots}, we show four contour plots of the CP asymmetry $\mathcal{A}$, in the plane $\delta$ and $\Gamma$, corresponding 
to four neutrino energies: 0.5, 2.7, 4.0 and 6.0 GeV. The latter values are representatives of the energy range of DUNE's neutrino beam
spectrum. We set the decoherence parameters as follows: $\Gamma=\Gamma_2=4\Gamma_4$. In each plot of this Figure, we have highlighted the 
iso-contours $\mathcal{A}$ produced by evaluating all the combinations between $\delta=\{\pi/2,3\pi/2\}$ and 
$\Gamma=\{0, 1, 2\}\times10^{-23}$ GeV. We do not display the iso-contour that contains $(3\pi/2, 0)$, given that this is practically a dot 
at this pair.

\begin{figure*}
\vspace{-4pt}
\centerline{
\includegraphics[width=0.5\textwidth]{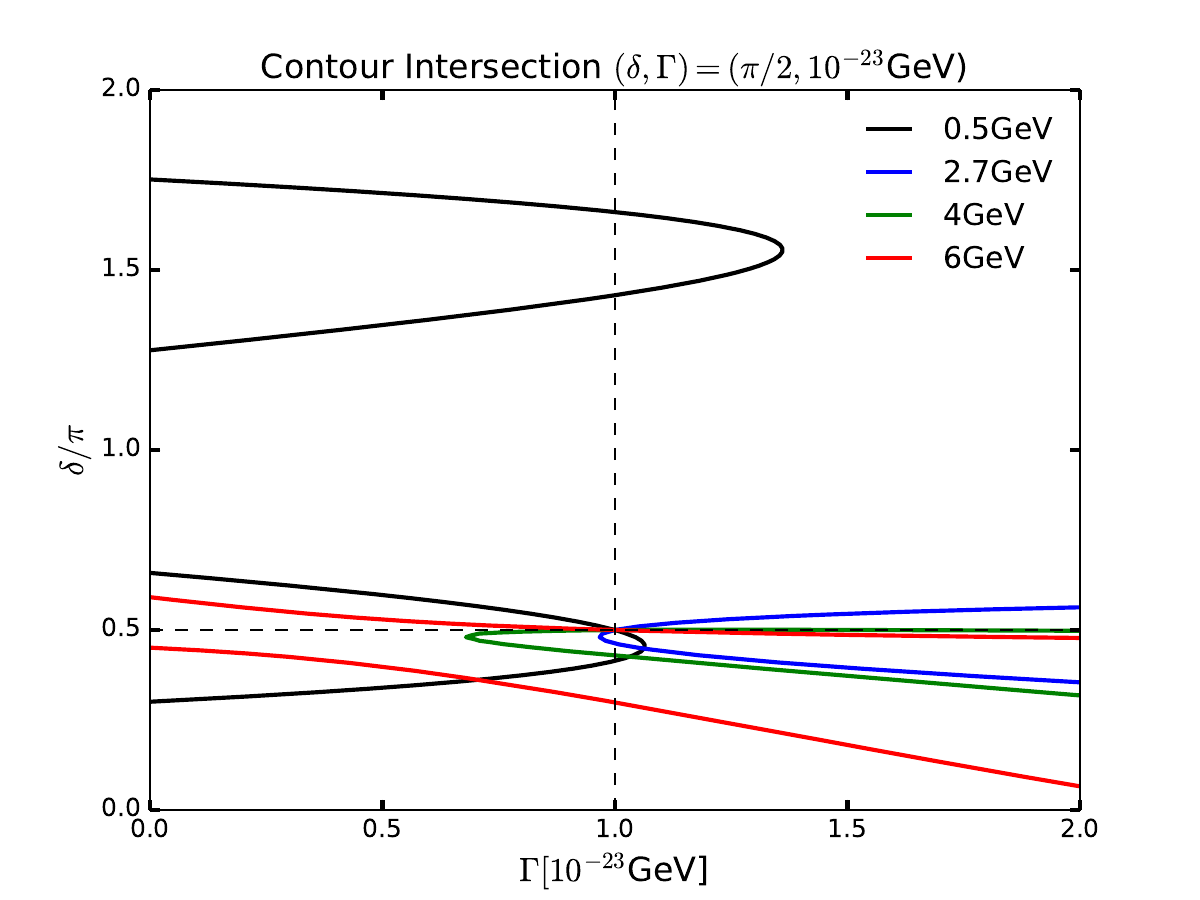}
\includegraphics[width=0.5\textwidth]{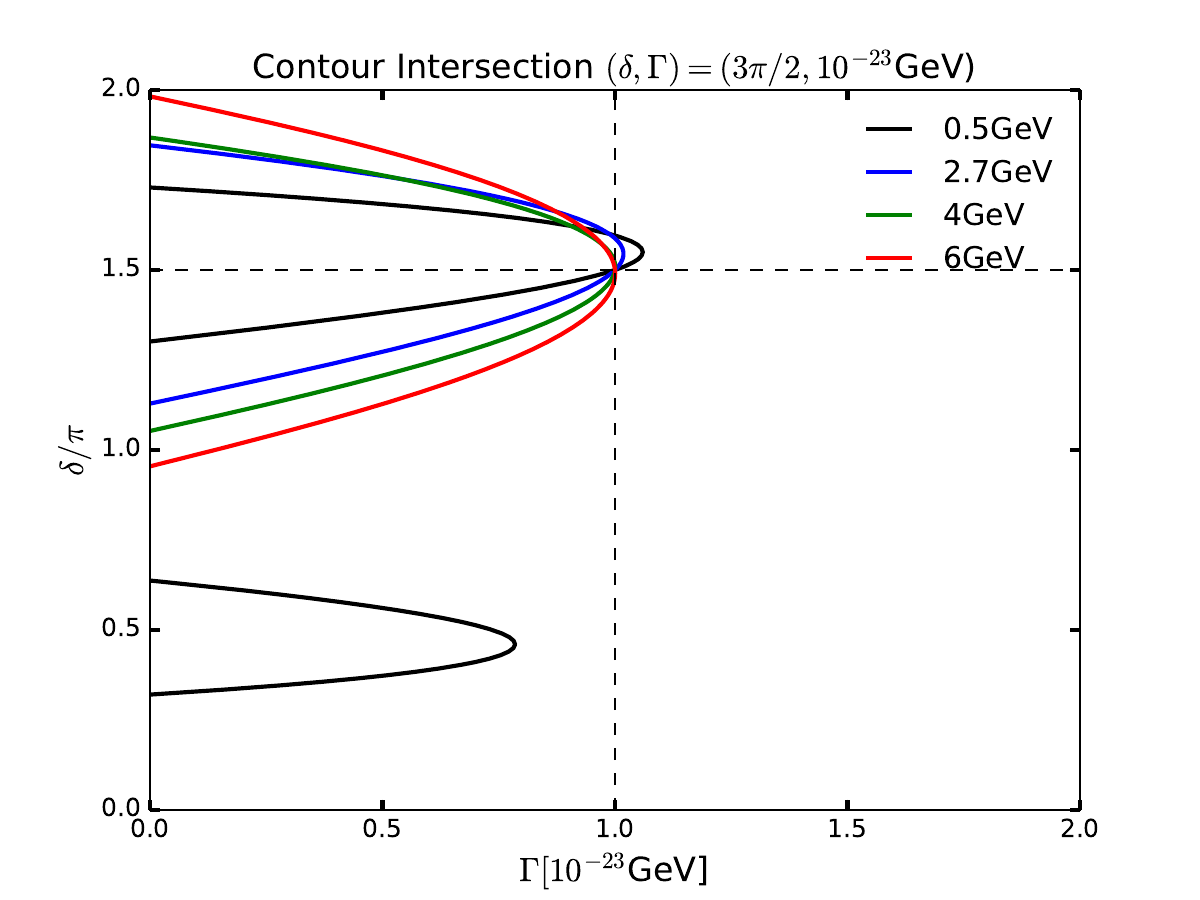}
}
\caption{Intersection of the iso-contours in Figure \ref{3NuCPPlots} for $(\delta=\Gamma)=(\pi/2,10^{-23}\text{GeV})$ (left panel) and 
$(\delta=\Gamma)=(3\pi/2,10^{-23}\text{GeV})$ (right panel) at the selected energies.}
\label{3NuCPIntersection}
\end{figure*}

First, we observe a common feature of the asymmetry $\mathcal{A}$ for these energies: as we increase the $\Gamma$, the asymmetry decreases 
with respect to its values at $\Gamma =0$ (the pure standard oscillation case), for any $\delta$. 
For example, looking at 2.7 GeV, we see  that for $\Gamma=0$ and $\delta=\pi/2 (3\pi/2)$, the asymmetry has values of 0.11(0.64). 
Meanwhile, for $\Gamma = 2\times10^{-23}$ and $\delta=\pi/2 (3\pi/2)$, the value of the asymmetry is 0.08(0.42), turning out in a decrement 
of the asymmetry by of 28\%(35\%) . Therefore, the tendency of this observable to vanish could be 
a signature of the decoherence phenomena. The degradation of the asymmetry can be explained by the term proportional to 
$\bar{\Gamma}_2 \theta_{13}$ in our approximate formula that was verified numerically.

Second, the iso-contour curves, which contain  points with the same CP asymmetry $\mathcal{A}$, reveal the existence of a degeneracy in the 
values of $(\delta,\Gamma)$. Taking, for instance, the 2.7 GeV curves, we appreciate that the iso-contour produced by the pair 
$(3\pi/2, 2\times10^{-23}\text{GeV})$ gives the same value of $\mathcal{A}$ as pairs like $(1.12\pi, 0)$ and $(1.84\pi,0)$, which refer to 
the pure standard oscillation case. The value of $\mathcal{A}$ at these pairs are lower than the one at ($3\pi/2$,0). 
Likewise, for $(\pi/2, 0)$ we note that the same value of $\mathcal{A}$ is obtained for $(0.6\pi, 2\times10^{-23}\text{GeV})$ and 
$(0.36\pi, 2\times10^{-23}\text{GeV})$. For this last case, the value of $\mathcal{A}$ is higher than the one at 
$(\pi/2, 2\times10^{-23}\text{GeV})$. 

Now we explore the possibility of lifting of these degeneracies for the following $(\delta, \Gamma)$ pairs: $(\pi/2,10^{-23}\text{GeV})$ 
and $(3\pi/2,10^{-23}\text{GeV})$, shown in Figure \ref{3NuCPIntersection}. In these Figures we displayed various iso-contours 
corresponding to different energies, 0.5 GeV, 2.7 GeV, 4.0 GeV and 6 GeV. We observe that for both $(\delta, \Gamma)$ pairs these four curves 
intersect only at the true point, thus solving the degeneracy. The clarity of the solution to the degeneracy problem in $(\delta, \Gamma)$ 
would be affected if we take consider a detailed experimental context. For the latter case we expect allowed regions around the true point 
and possibly around others (at some confidence level), being difficult to predict at what confidence level we may identify the true solution. 
An implication of this is the possibility of imposing upper limits in $\Gamma$. 


Although it is not shown, we also looked at $\Delta P= P(\nu_\mu\to\nu_e)-P(\bar{\nu}_\mu\to\bar{\nu}_e)$. Here the decrement is 13\% (16\%) 
for values of $\Gamma =2\times 10^{-23}$ GeV and $\delta=\pi/2 (3\pi/2)$.
In the asymmetry $\mathcal{A}$ these differences are magnified due to the increment in the denominator when decoherence is present. 

In the latest global analysis of neutrino data, using the standard oscillation hypothesis, a hint for $\delta=1.3\pi$ has been found 
\cite{Nufit}. If decoherence is present as a subleading effect, the latter result has to be taken with care, due to the degeneracy 
introduced by decoherence in the measurement of $\delta$. In fact, it would be valuable to assess this distortion in the context 
of a simulation that convolutes the neutrino probabilities with the cross-section, fluxes, efficiencies, resolutions, etc, in facilities 
such as DUNE \cite{Deco2}.

\section{SUMMARY AND CONCLUSIONS}
We have re-examined neutrino oscillation probabilities in matter in the presence of subleading decoherence effects. The effect of rotating 
from the VMB to the MMB was mentioned, pointing out that such a rotation inevitably changes the form of the decoherence matrix
in the new basis. The inability of substituting oscillation parameters in the vacuum oscillation formula with their effective values in matter
for arbitrary decoherence matrices has been heavily emphasized, providing strict conditions which must be satisfied for this method
to be viable. In the context of three generation mixing, we have presented a perturbative approach to the decoherence problem, valid for 
$\Gamma<10^{-23}$ GeV for the DUNE  baseline, which explains the prominent features of the oscillation probabilities. More importantly,
we show that a term proportional to $\Gamma\theta_{13}$ connects decoherence with CP violation. This term causes a $(\delta,\Gamma)$ 
degeneracy associated with a reduction of the CP asymmetry, when the decoherence parameter increases. 
We have shown that it is possible to lift these degeneracies, at the level of neutrino oscillation probabilities, combining a set of energies 
from the DUNE neutrino energy beam. However, to give a more precise answer, a realistic analysis is needed. Without a doubt, a 
future measurement of the CP asymmetry is going to be a useful tool for either bring to light the decoherence phenomena or constrain it. 

\section{ACKNOWLEDGEMENTS}
The authors acknowledge funding by the {\it Direcci\'on de Gesti\'on de la Investigaci\'on} at PUCP, through grant DGI-2015-3-0026 and DGI-2017-3-0019. They would also like to thank J.~Jones-P\'erez and C.~Arg\"uelles for a very careful reading of the manuscript and useful suggestions. 

\pagebreak
\appendix
\section{Effective mixing angles and masses}
In Reference \cite{Freund01}, $\tilde{m}^2_{ij}$ and $\tilde{\theta}_{ij}$ were expressed as functions
of the small parameters $\alpha = \Delta m^2_{21}/\Delta m^2_{31}$ and $\sin\theta_{13}$. We follow a similar approach,
difference being that we rewrite the effective quantities as power series in $\alpha,\theta_{13}$ up to second order
\begin{eqnarray}\nonumber
\sin\tilde{\theta}_{13}&= &\frac{1}{1-A}\theta_{13}+\alpha\theta_{13} A\sin^22\theta_{12}\\\nonumber
\sin\tilde{\theta}_{23}&= &\sin\theta_{23}+\alpha\theta_{13}\frac{A \cos\delta \cos\theta_{23}\sin2\theta_{12}}{2(1-A)}\\\nonumber
\sin\tilde{\theta}_{12}&= &-\alpha\frac{\cos\theta_{12}\sin\theta_{12}}{A}-\alpha^2 \frac{\sin4\theta_{12}}{4A^2}-\\\nonumber
&\quad&\alpha\theta_{13}\frac{\sin2\theta_{12}}{2A}\\
\sin\tilde{\delta}& = &\sin\delta
\label{EffectiveAngleFormula}
\end{eqnarray}
which facilitates the calculations presented in our method.

Similarly, the effective masses are expanded as follows
\begin{eqnarray}\nonumber
\tilde{\Delta}_{21}&=&\Delta_{31} \left(-A+\alpha\cos2\theta_{12}+\frac{A\theta_{13}^2}{1-A}\right)\\\nonumber
\tilde{\Delta}_{32}&=&\Delta_{31} \left(1-\alpha\cos^2\theta_{12}+\frac{\alpha^2\sin2\theta_{12}}{2A}+\frac{A\theta_{13}^2}{2(1-A)}\right)\\
\tilde{\Delta}_{31}&=&\Delta_{31} \left(1-A -\frac{\alpha\sin^2\theta_{12}}{1-A}\right)
\label{EffectiveMassFormula}
\end{eqnarray}

\section{Decoherence matrices invariant to MMB rotations for three generations}
\label{MMBrotations_analytical}
Suppose we wish to find a matrix that remains invariant after performing rotations of the form \eqref{NRotationFormula}. As an example, 
we choose the matrix $M_D^V=-\text{Diag}(\Gamma_1,\Gamma_2,\dots\Gamma_{8})$. We will work in a power series
of $\alpha,\theta_{13}$ as outlined in \ref{3NuDecoherence}, using the approximate formulas for the effective angles and masses. 
The rotated matrix $M_D^M$ is expressed as a power series as shown in Eq.\eqref{DecoMatrixDecomposition}. If we demand $M_D^M=M_D^V$, and a diagonal form for these matrices,
it follows that $M_\Gamma,M_{\Gamma\alpha},M_{\Gamma\theta},\dots$ are all diagonal. Starting off with the leading term, we have
\begin{equation}
M_\Gamma = -\begin{pmatrix}
X & 0 & P_1 & 0&0&0&0&0\\
0 & \Gamma_2 & 0&0&0&0&0&0\\
P_1& 0 & X & 0&0&0&0&0\\
0&0&0&X & 0&P_2&0&0\\
0&0&0&0&X&0&P_3&0\\
0&0&0&P_2&0&X&0&0\\
0&0&0&0&P_3&0&X&0\\
0&0&0&0&0&0&0&\Gamma_8
\end{pmatrix}
\end{equation}
where
\begin{eqnarray}\nonumber
P_1 = \frac{1}{2}\sin(4\theta_{12})(\Gamma_3-\Gamma_1)\\\nonumber
P_2 = \frac{1}{2}\sin(2\theta_{12})(\Gamma_6-\Gamma_4)\\
P_3 = \frac{1}{2}\sin(2\theta_{12})(\Gamma_7-\Gamma_5)
\label{off1stord}
\end{eqnarray}
and $X$ are expressions that we are not currently interested in, since they belong to the main diagonal. If $M_\Gamma$ is diagonal, 
it follows that $\Gamma_1=\Gamma_3,\Gamma_4=\Gamma_6$ and $\Gamma_5=\Gamma_7$. Imposing this condition on $M_{\Gamma\theta}$, we find that
\begin{equation}
M_{\Gamma\theta}=\frac{A\theta_{13}}{A-1}\begin{pmatrix}
0&0&0&0&0&Q_1&Q_2&0\\
0&0&0&0&0&Q_3&Q_4&0\\
0&0&0&Q_1&Q_2&0&0&0\\
0&0&Q_1&0&0&0&0&Q_5\\
0&0&Q_2&0&0&0&0&Q_6\\
Q_1&Q_3&0&0&0&0&0&0\\
Q_2&Q_4&0&0&0&0&0&0\\
0&0&0&Q_5&Q_6&0&0&0
\end{pmatrix}
\end{equation}
with
\begin{eqnarray}\nonumber
Q_1& = (\Gamma_1-\Gamma_4)\cos\delta\\\nonumber
Q_2& = (\Gamma_1-\Gamma_5)\sin\delta\\\nonumber
Q_3& = (\Gamma_2-\Gamma_4)\sin\delta\\\nonumber
Q_4&= (\Gamma_5-\Gamma_2)\cos\delta\\\nonumber
Q_5&= \sqrt{3}(\Gamma_8-\Gamma_3)\cos\delta\\
Q_5&= \sqrt{3}(\Gamma_8-\Gamma_5)\sin\delta
\label{off2ndord}
\end{eqnarray}
For this matrix to be diagonal, $M_{\Gamma\theta}$ must vanish, 
this turns out into three possibilities: 
\begin{enumerate}
\item All $\Gamma_i$ are equal and $M_D^M$ (or $M_D^V$) are
proportional to the identity.
\item For the case $\delta=0,\pi$, we have $\Gamma_1=\Gamma_4$,  $\Gamma_3=\Gamma_8$ and $\Gamma_2=\Gamma_5$. 
\item For the case $\delta=\pi/2, 3\pi/2$, we have $\Gamma_1=\Gamma_5$, $\Gamma_2=\Gamma_4$ and $\Gamma_5=\Gamma_8$.
\end{enumerate}
plus the extra conditions given for a diagonal $M_\Gamma$, relevant for the second and third case. The explicit expressions for $M_D^M$ (or $M_D^V$) are given in Eqs. \eqref{diagvalid0} and \eqref{diagvalidpi2} for $\delta=0,\pi$
and  $\delta=\pi/2, 3\pi/2$, respectively. 

\break

\end{document}